\documentclass[twocolumn,english,aps,prd,nofootinbib]{revtex4}
\usepackage{lmodern}
\usepackage[T1]{fontenc}
\usepackage[latin9]{inputenc}
\usepackage{xcolor}
\usepackage{babel}
\usepackage{amsmath}
\usepackage{amssymb}
\usepackage{graphicx}
\usepackage[unicode=true]
{hyperref}

\makeatletter
\@ifundefined{textcolor}{}
{%
\definecolor{BLACK}{gray}{0}
\definecolor{WHITE}{gray}{1}
\definecolor{RED}{rgb}{1,0,0}
\definecolor{GREEN}{rgb}{0,1,0}
\definecolor{BLUE}{rgb}{0,0,1}
\definecolor{CYAN}{cmyk}{1,0,0,0}
\definecolor{MAGENTA}{cmyk}{0,1,0,0}
\definecolor{YELLOW}{cmyk}{0,0,1,0}
}

\usepackage{enumerate}

\makeatother

\begin{document}
\title{A stationary axisymmetric vacuum solution for pure $R^{2}$ gravity}
\author{Mustapha Azreg-A\"inou}
\email{azreg@baskent.edu.tr}

\affiliation{Ba\c{s}kent University, Engineering Faculty, Ba\u{g}l{\i}ca Campus,
06790-Ankara, Turkey}
\author{Hoang Ky Nguyen}
\email{hoang.nguyen@ubbcluj.ro}

\affiliation{Department of Physics, Babe\c{s}--Bolyai University, 400084 Cluj-Napoca,
Romania}
\begin{abstract}
The closed-form expression for pure $\mathcal{R}^{2}$ vacuum solution obtained in Phys. Rev. D \textbf{107}, 104008 (2023) lends itself to a generalization to axisymmetric setup via the modified Newman--Janis algorithm. We adopt the procedure put forth in Phys. Rev. D \textbf{90}, 064041 (2014) bypassing the complexification of the radial coordinate. The procedure presumes the existence of Boyer-Lindquist coordinates. Using the Event Horizon Telescope Collaboration results, we model the central black hole M87{*} by the thus obtained exact rotating metric, depending on the mass, rotation parameter and a third dimensionless parameter. The latter is constrained upon investigating the shadow angular size assuming mass and rotation parameters are those of M87{*}. Stability is investigated.
\end{abstract}
\maketitle

\section{Introduction}

The family of $f(\mathcal{R})$ theories is an active arena of research
in modified gravity. It is considerably less involved than theories
containing the Ricci or Riemann tensors in their action \cite{Nojiri-2011,Clifton-2011,deFelice-2010,Sotiriou-2008}.
It is also advantaged by being ghost-free, as its scalar degree of
freedom, when moving from the Jordan frame to the Einstein frame,
involves derivatives no higher than two. This theory offers a number
of exact analytical solutions, some of which project non-constant
scalar curvature \cite{Sebastiani-2010,Capozziello-2007,Gurses-2012,Pravda-2017}.
However, these solutions do not align with the Schwarzschild metric
and fail to pass tests in the Solar System and binary stars.\vskip4pt

Prior to proposing $f(R)$ gravity \cite{Buchdahl-1970},
Buchdahl considered the pure $R^{2}$ action around 1962, which is
a simple and natural extension of General Relativity \cite{Buchdahl-1962}.
Its action contains one single term in the gravitation sector, $\frac{1}{2\kappa}\int d^{4}x\sqrt{-g}\,\mathcal{R}^{2}$,
making it more parsimonious than Brans-Dicke gravity which involves
the Brans-Dicke parameter $\omega$, and conformal gravity which involves
the nuanced Bach tensor. The action is scale invariant, with the parameter
$\kappa$ being dimensionless. In vacuum, the action leads to the
field equation 
\begin{equation}
	\mathcal{R}\left(\mathcal{R}_{\mu\nu}-\frac{1}{4}g_{\mu\nu}\mathcal{R}\right)+\left(g_{\mu\nu}\square-\nabla_{\mu}\nabla_{\nu}\right)\mathcal{R}=0,\label{eq:R2-field-eqn}
\end{equation}
and, consequently, the trace equation
\begin{equation}
	\square\,\mathcal{R}=0.\label{eq:trace-eqn}
\end{equation}
In \cite{Buchdahl-1962} Buchdahl demonstrated that the trace equation
\eqref{eq:trace-eqn} allows for a \emph{non-constant}{}
Ricci scalar field. To establish this, he expressed the metric in
the ``harmonic gauge'' as
\begin{equation}
	ds^{2}=-A(u)dt^{2}+B(u)du^{2}+\sqrt{B(u)/A(u)}d\Omega^{2},\label{eq:harmonic-ds}
\end{equation}
where $A(u)$ and $B(u)$ are two unknown functions of $u$, a harmonic
coordinate in the radial direction. For this metric, the d'Alembertian
acting on a scalar field is
\begin{align}
	\square\,\mathcal{R} & \equiv\frac{1}{\sqrt{-g}}\partial_{\mu}\left(\sqrt{-g}g^{\mu\nu}\partial_{\nu}\mathcal{R}\right)=B^{-1}\frac{d^{2}\mathcal{R}}{du^{2}}.\label{eq:dAlembertian}
\end{align}
The trace equation in vacuum \eqref{eq:trace-eqn} directly yields
$\frac{d^{2}\mathcal{R}}{du^{2}}=0$, resulting in $\mathcal{R}$
being an affine function of $u$, viz.
\begin{equation}
	\mathcal{R}=\Lambda+k\,u,\label{eq:affine}
\end{equation}
where $\Lambda$ and $k$ are two constants.\vskip8pt

Recently, one of us revisited Buchdahl's 1962 work
and discovered a class of spacetime solutions for pure $\mathcal{R}^{2}$
gravity \cite{Nguyen-2022-Buchdahl}. Consistent with the finding
expressed in Eq. \eqref{eq:affine}, these Buchdahl-inspired solutions
are asymptotically de Sitter and exhibit \emph{non-constant}
scalar curvature. In addition to the asymptotic value of the scalar
curvature at spatial infinity, the solutions are specified by a new
(Buchdahl) parameter, $k$, which has dimensions of length and is
of higher-derivative nature. \vskip4pt

\emph{The role of the Buchdahl parameter $k$ can
	be likened to the ``scalar charge'' in scalar-tensor theories}.
As demonstrated in \cite{Bronnikov-1998}, for the BD action $L_{\text{BD}}=\sqrt{-g}\Bigl[\phi\,\mathcal{R}+\frac{\omega}{\phi}\phi_{;\mu}\phi^{;\mu}\Bigr]$,
by switching from the Jordan frame to the Einstein frame via the
transformations: 
\begin{equation}
	\tilde{g}_{\mu\nu}:=\phi\,g_{\mu\nu}\ \ \ \text{and}\ \ \ \varphi:=\sqrt{|\omega+3/2|}\,\ln\phi \,,
\end{equation}
the action takes the form of General Relativity with a minimally coupled
scalar field $\varphi$, per
\begin{equation}
	L_{\text{BD}}=\sqrt{-\tilde{g}}\Bigl[\tilde{\mathcal{R}}+\varepsilon\,\tilde{g}^{\mu\nu}\partial_{\mu}\varphi\partial_{\nu}\varphi\Bigr];\ \ \varepsilon:=\text{sgn}\Bigl(\omega+\frac{3}{2}\Bigr).
\end{equation}
The resulting field equations are
\begin{align}
	\tilde{\mathcal{R}}_{\text{\textmu}\nu} & =-\varepsilon\,\partial_{\mu}\varphi\partial_{\nu}\varphi\,,\label{eq:BD-Einstein-g}
\end{align}
and
\begin{equation}
	\tilde{\square}\,\varphi=0.\label{eq:BD-Einstein-varphi}
\end{equation}
Similarly to the steps taken in Eqs.$\,$\eqref{eq:harmonic-ds}--\eqref{eq:affine},
in the harmonic coordinate $u$, the harmonic equation \eqref{eq:BD-Einstein-varphi}
directly yields
\begin{equation}
	\varphi=\varphi_{0}+C\,u\,;\ \ \ \ \ \ \varphi_{0},\,C=\text{const}\,,\label{eq:phi-affine}
\end{equation}
with $C$ representing a {\emph{scalar charge}}, a notion first advanced by Bronnikov in 1973 \cite{Bronnikov-1973}. When $C=0$, the Schwarzschild metric is recovered. In general, $C\in\mathbb{R}$. The scalar charge plays
a central role in generating new physics for Brans-Dicke gravity \cite{Bronnikov-1998}.
For $C\neq0$, as $\varphi$ varies, resulting in $\tilde{\mathcal{R}}_{\mu\nu}\neq0$
in vacuum per Eq. \eqref{eq:BD-Einstein-g}, non-Schwarzschild solutions
are obtained. Furthermore, it has been determined that these solutions
in scalar-tensor theories are typically stable \cite{Bronnikov-1998,Matsuda-1972,Campanelli-1993}.\vskip4pt

Considering the close analogy of the two harmonic
equations, Eq. \eqref{eq:affine} vs Eq. \eqref{eq:phi-affine}, and
of the pure $\mathcal{R}^{2}$ and Brans-Dicke theories \footnote{We also establish that a special case of the Buchdahl-inspired
solutions is related to the Campanelli-Lousto solution of Brans-Dicke gravity; see our comment at the end of Section \ref{sec:Buchdahl-inspired-solutions}. This connection further underscores the similarities and shared characteristics
between these two gravitational theories.}, it follows that the Buchdahl parameter $k$ can be considered as the \emph{scalar charge} and should take on any value in $\mathbb{R}$ unrestricted. \vskip4pt

Within the class of Buchdahl-inspired solutions mentioned
above, a specific case has been derived in Ref. Ref.~\cite{Nguyen-2022-Lambda0}
which provides an exact closed analytical solution describing asymptotically
flat spacetimes. This special Buchdahl-inspired metric recovers Schwarzschild
when the Buchdahl parameter $k$ vanishes. Yet, for $k\neq0$, this
solution exhibits novel intriguing properties for $\mathcal{R}^{2}$
spacetimes \cite{Nguyen-2022-Lambda0}. Pure $\mathcal{R}^{2}$ gravity is thus an example of a higher-order theory capable of producing a diverse range of phenomena, even in the absence of complex ingredients such as torsion, non-metricity, or non-locality.\vskip4pt

The new metrics invite extensions to the stationary axisymmetric setup,
via the use of the Newman-Janis algorithm (NJA). The method starts
with a ``seed'' static spherically symmetric metric in a closed form
\begin{equation}
ds_{\text{seed}}^{2}=-G(r)dt^{2}+\frac{dr^{2}}{F(r)}+H(r)\left(d\theta^{2}+\sin^{2}\theta d\varphi^{2}\right).\label{eq:seed}
\end{equation}
A crucial step in the NJA is the complexification of the radial coordinate,
viz. $r\rightarrow r+ia\cos\theta$, in which the following replacements
are adopted: 
\begin{align}
r^{2} & \rightarrow(r+ia\cos\theta)(r-ia\cos\theta)=r^{2}+a^{2}\cos^{2}\theta,\label{eq:complexification}\\
\frac{1}{r} & \rightarrow\frac{1}{2}\left(\frac{1}{r+ia\cos\theta}+\frac{1}{r-ia\cos\theta}\right)=\frac{r}{r^{2}+a^{2}\cos^{2}\theta}.\nonumber 
\end{align}
For the ``seed'' Reissner--Nordstr\"om metric, viz. $F(r)=G(r)=1-\frac{2M}{r}+\frac{Q^{2}}{r^{2}}$
and $H(r)=r^{2}$, the NJA aptly produces the Kerr-Newman metric that
describes a charged rotating black hole in General Relativity. \vskip4pt

As the complexification scheme adopted in Eq. \eqref{eq:complexification}
is rather ad hoc, in Refs.$\ $\cite{Azreg-Ainou-B,Azreg-Ainou-C,Azreg-Ainou-2014}, the other
author of us proposed another route with a higher degree of plausibility.
It assumes the existence of Boyer-Lindquist coordinates and imposes
an integrability condition in some of its coordinate transformations.
The procedure is an improvement over the NJA and bypasses the complexification
step. \vskip4pt

The purpose of our current paper is to apply the non-complexification
technique put forth in Ref.$\ $\cite{Azreg-Ainou-2014} to the Buchdahl-inspired
solutions, in order to obtain rotating solutions for uncharged rotating
sources in pure $\mathcal{R}^{2}$ gravity. We will be able to derive an exact rotating solution up to a conformal factor, the determination of which requires solving some challenging partial differential equation. However, for the purpose of this work, the exact form of the conformal factor is not needed and will not affect our conclusions. Our paper is organized as follows. In Section~\ref{sec:Buchdahl-inspired-solutions}
we first review the general and special Buchdahl-inspired metrics
obtained in \cite{Nguyen-2022-Buchdahl,Nguyen-2022-Lambda0}. In
Sections~\ref{sec:A-coordinate-change} and~\ref{sec:Moving-to-Einstein-frame},
we recast the generic metric in a new set of coordinates to bring
it to a ``Schwarzschild gauge'' in the Einstein frame. We then apply
the non-complexification algorithm to this ``seed'' metric in Section~\ref{sec:Non-complexification} and to the special case where $\Lambda=0$
in Section~\ref{sec:Case-of-Lambda0}. Finally, in Section~\ref{sec:Shadow}
we apply the rotation metric to the M87{*} shadow and obtain
a bound for the Buchdahl parameter $k$. Section~\ref{sec:Stability} is devoted to the stability analysis. We conclude in Section~\ref{sec:Summary}.

\section{\label{sec:Buchdahl-inspired-solutions}Buchdahl-inspired solutions}

In Ref.$\ $\cite{Nguyen-2022-Buchdahl} the static spherically symmetric
vacuo solution to the $\mathcal{R}^{2}$ field equation \eqref{eq:R2-field-eqn}
was found to be expressible in terms of two auxiliary functions $p(r)$
and $q(r)$ \small
\begin{equation}
ds^{2}=e^{k\int\frac{dr}{r\,q(r)}}\left\{ -\frac{p(r)q(r)}{r}dt^{2}+\frac{p(r)\,r}{q(r)}dr^{2}+r^{2}d\Omega^{2}\right\}, \label{eq:B-metric}
\end{equation}
\normalsize in which $p(r)$ and $q(r)$ obey a coupled set of first-order
``evolution'' type ordinary differential equations (ODE): 
\begin{align}
\frac{dp(r)}{dr} & =\frac{3\,k^{2}}{4\,r}\frac{p(r)}{q^{2}(r)},\label{eq:evol-p}\\
\frac{dq(r)}{dr} & =(1-\Lambda\,r^{2})\,p(r),\label{eq:evol-q}
\end{align}
and the Ricci scalar is given by
\begin{equation}
\mathcal{R}(r)=4\Lambda\,e^{-k\int\frac{dr}{r\,q(r)}}.\label{eq:Ricci-B}
\end{equation}

The generic Buchdahl-inspire metric is specified by four parameters,
viz. $\left\{ \Lambda,k,p(r_{0}),q(r_{0})\right\} $, reflecting the
fourth order nature of the $\mathcal{R}^{2}$ action. The Ricci scalar approaches
$4\Lambda$ at spatial infinity, indicating its asymptotic de Sitter
behavior. The new (Buchdahl) parameter $k$, which has units of length,
enables the metric to deviate from the Schwarzschild-de Sitter metric.
In the limit where $k=0$, the metric recovers the Schwarzschild-de
Sitter solution.\vskip4pt

For the case where $\Lambda=0$, the ``evolution'' ODE's \eqref{eq:evol-p}
and \eqref{eq:evol-q} are fully \emph{soluble}. The solution is \cite{Nguyen-2022-Lambda0}
\begin{align}
r & =\left|q-q_{+}\right|^{\frac{q_{+}}{q_{+}-q_{-}}}\left|q-q_{-}\right|^{-\frac{q_{-}}{q_{+}-q_{-}}},\label{eq:r-vs-q}\\
p & =\frac{(q-q_{+})(q-q_{-})}{r\,q},\\
q_{\pm} & :=\frac{1}{2}\Bigl(-r_{\text{s}}\pm\sqrt{r_{\text{s}}^{2}+3k^{2}}\Bigr),
\end{align}
with $r_{\text{s}}$ being an integration constant and $q_{\pm}$
the two real roots of the algebraic equation $q^{2}+r_{\text{s}}q-\frac{3k^{2}}{4}=0$.
Interestingly, although the Buchdahl ODE \eqref{eq:evol-q} involves
$q$ as a function of $r$, the solution \eqref{eq:r-vs-q} has $r$
expressed in terms of $q$.\vskip4pt

Ref.~\cite{Nguyen-2022-Lambda0} made a further transformation
from $r$ to a \emph{new} radial coordinate $\rho$ per{\small{}
\begin{align}
	r(\rho) & :=\frac{\zeta\,r_{\text{s}}\left|1-\frac{r_{\text{s}}}{\rho}\right|^{\frac{1}{2}(\zeta-1)}}{1-\text{sgn}\bigl(1-\frac{r_{\text{s}}}{\rho}\bigr)\left|1-\frac{r_{\text{s}}}{\rho}\right|^{\zeta}},\label{eq:r-vs-rho}\\
	\zeta & :=\sqrt{1+\frac{3k^{2}}{r_{\text{s}}^{2}}}.
\end{align}}%
In this new coordinate, the special Buchdahl-inspired metric takes
on a strikingly well-structured form. It is specified by a ``Schwarzschild''
radius $r_{\text{s}}$ and \emph{the scaled (dimensionless) Buchdahl
parameter} $\tilde{k}:=k/r_{\text{s}}$ per {\small
\begin{equation}
ds^{2}=\left|1-\frac{r_{\text{s}}}{\rho}\right|^{\tilde{k}}\Big\{-\Bigl(1-\frac{r_{\text{s}}}{\rho}\Bigr)dt^{2}+\frac{d\rho^{2}}{1-\frac{r_{\text{s}}}{\rho}}\frac{r^{4}(\rho)}{\rho^{4}}+r^{2}(\rho)\,d\Omega^{2}\Big\}.\label{eq:special-B-metric}
\end{equation}}%
For $\tilde{k}\neq0$, the special Buchdahl-inspired
metric is not Ricci-flat and is therefore non-Schwarzschild. However,
it is asymptotically flat and Ricci-scalar-flat. When $\tilde{k}=0$,
it precisely recovers the Schwarzschild metric since $r(\rho)\equiv\rho$
for all $\rho\in\mathbb{R}$, hence being viable of passing tests
in the Solar System and binary star systems.\vskip4pt

The metric given in Eq.~\eqref{eq:special-B-metric}
possesses two additional noteworthy properties:\vskip4pt

(I) \emph{Non-triviality of the
	asymptotically flat Buchdahl-inspired metric: }The
metric in \eqref{eq:special-B-metric} is Ricci-scalar flat, viz.
$\mathcal{R}\equiv0\ \ \forall r\in\mathbb{R}^{+}$. While it is known
that any null-Ricci-scalar metric is automatically a vacuum solution
to the pure $\mathcal{R}^{2}$ field equation (thus forming a trivial
branch of solutions), the metric in \eqref{eq:special-B-metric} is
non-trivial since it satisfies a ``stronger'' version the field
equation:
\begin{equation}
	\mathcal{R}_{\mu\nu}-\frac{1}{4}g_{\mu\nu}\mathcal{R}+g_{\mu\nu}\mathcal{R}^{-1}\square\,\mathcal{R}-\mathcal{R}^{-1}\nabla_{\mu}\nabla_{\nu}\mathcal{R}=0.\label{eq:R2-field-eqn-stronger}
\end{equation}
Despite $\mathcal{R}^{-1}$ being singular, the combinations $\mathcal{R}^{-1}\square\,\mathcal{R}$
and $\mathcal{R}^{-1}\nabla_{\mu}\nabla_{\nu}\mathcal{R}$ are free
of singularities. This result has recently been reported in \cite{Nguyen-2023-Nontrivial}.\vskip4pt

(II) \emph{Embedding into the generalized
	Campanelli-Lousto solution of Brans-Dicke gravity:} Furthermore, it has been shown in \cite{WH-2023,WEC-2023}
that the metric in \eqref{eq:special-B-metric} is equivalent to
\begin{align}
	ds^{2} & =-\text{s}\left|1-\frac{\zeta r_{\text{s}}}{r}\right|^{A}dt^{2}+\text{s}\left|1-\frac{\zeta r_{\text{s}}}{r}\right|^{B}dr^{2}\nonumber \\
	& \ \ \ \ \ \ \ \ \ \ +\left|1-\frac{\zeta r_{\text{s}}}{r}\right|^{B+1}r^{2}d\Omega^{2},\label{eq:CL}
\end{align}
in which $\text{s}:=\text{sgn}\left(1-\frac{\zeta r_{\text{s}}}{r}\right)$
is the signum function, and
\begin{equation}
	A:=\frac{\tilde{k}+1}{\zeta},\ \ \ \ \ B:=\frac{\tilde{k}-1}{\zeta}.
\end{equation}
Expression \eqref{eq:CL} is a special case of the \emph{generalized}
Campanelli-Lousto solution that we reported for Brans-Dicke gravity
\cite{WEC-2023}. This relationship between the two solutions further
highlights the shared characteristics of pure $\mathcal{R}^{2}$ gravity
and Brans-Dicke gravity.

\section{\label{sec:A-coordinate-change}A coordinate change for Buchdahl-inspired
solution}

Regarding the metric in \eqref{eq:B-metric}, let us make a coordinate
change $r(R)$ to make $-g_{tt}$ and $g_{RR}$, when excluding the
conformal factor, reciprocal to each other. The change of coordinate
is meant to be
\begin{equation}
\left\{ \frac{p(r)q(r)}{r}\right\} \times\left\{ \frac{p(r)\,r}{q(r)}\frac{dr^{2}}{dR^{2}}\right\} =1,
\end{equation}
hence yielding 
\begin{equation}
dR=p(r)\,dr.
\end{equation}
In this new coordinate, the evolution rules \eqref{eq:evol-p}--\eqref{eq:evol-q}
are enlarged to three coupled equations: 
\begin{align}
\frac{dp(R)}{dR} & =\frac{3\,k^{2}}{4\,r(R)}\frac{1}{q^{2}(R)},\label{eq:evol-1}\\
\frac{dq(R)}{dR} & =1-\Lambda\,r^{2}(R),\label{eq:evol-2}\\
\frac{dr(R)}{dR} & =\frac{1}{p(R)}.\label{eq:evol-3}
\end{align}
Note that the right hand sides of these equations do not depend explicitly on
$R$; they are thus ``translationally invariant'' with respect to
arbitrary shift $R\rightarrow R+R_{0}$. The integral in the conformal
factor of \eqref{eq:B-metric} becomes 
\begin{equation}
\int\frac{dr}{r\,q(r)}=\int\frac{dR}{p(R)q(R)r(R)}.
\end{equation}
Despite the appearance of three variables $\{p,q,r\}$, the metric
in the new coordinate $R$ still involves only two degrees of freedom
(as a general result of gauge choice in the static spherically symmetric
setup), viz. $r(R)$ and a function $\Psi(R)$ defined as 
\begin{equation}
\Psi(R):=\frac{p(R)q(R)}{r(R)},\label{Psi-def}
\end{equation}
namely \small
\begin{equation}
ds^{2}=e^{k\int\frac{dR}{\Psi(R)r^{2}(R)}}\biggl\{-\Psi(R)dt^{2}+\frac{dR^{2}}{\Psi(R)}+r^{2}(R)d\Omega^{2}\biggr\}.\label{eq:B0}
\end{equation}
\normalsize The Ricci scalar is
\begin{equation}
\mathcal{R}(R)=4\Lambda\,e^{-k\int\frac{dR}{\Psi(R)r^{2}(R)}}.\label{eq:Ricci-new}
\end{equation}
Note that one might be tempted to start from Eq.~\eqref{eq:B0} then
derive two ODE's for $\Psi(R)$ and $r(R)$ but the resulting ODE's
would be non-linear and high-order. The success of Buchdahl's program
is to decompose $\Psi(R)$ into $\{p,q,r\}$ that obey ``simpler''
coupled ODE's, Eqs. \eqref{eq:evol-1}--\eqref{eq:evol-3}.\vskip4pt

The purpose of this section was to bring metric~\eqref{eq:B-metric} to form~\eqref{eq:B0}, which upon moving to the Einstein frame (Sec.~\ref{sec:Moving-to-Einstein-frame}) will result in a ``seed'' metric suitable for generating its rotating counterpart, as will be done in Sec.~\ref{sec:Non-complexification}.

\section{\label{sec:Moving-to-Einstein-frame}Moving to Einstein frame}

Metric \eqref{eq:B0}--\eqref{eq:Ricci-new}\small
\begin{equation}
g_{\mu\nu}dx^{\mu}dx^{\nu}=\frac{4\Lambda}{\mathcal{R}(R)}\left\{ -\Psi(R)dt^{2}+\frac{dR^{2}}{\Psi(R)}+r^{2}(R)d\Omega^{2}\right\}, \label{eq:B1}
\end{equation}
\normalsize is the vacuo solution to the pure $\mathcal{R}^{2}$
action in the Jordan frame
\begin{equation}
\int d^{4}x\sqrt{-g}\frac{1}{2\kappa}\mathcal{R}^{2}.
\end{equation}
The action can be recast in terms of an auxiliary scalar field $\omega$
as 
\begin{equation}
\int d^{4}x\sqrt{-g}\,\frac{4\Lambda}{\kappa}\left[\omega\mathcal{R}-2\Lambda\,\omega^{2}\right].\label{eq:action-Einstein-frame}
\end{equation}
Upon a conformal transformation 
\begin{equation}
g_{\mu\nu}=\omega^{-1}\,\tilde{g}_{\mu\nu}\,,
\end{equation}
the term $\sqrt{-g}\,\omega\mathcal{R}$ becomes (e.g., see Formula
(27) in \cite{Carneiro-2004}) \small
\begin{equation}
\sqrt{-g}\,\omega\mathcal{R}=\left(\omega^{-2}\sqrt{-\tilde{g}}\right)\omega\,\omega\left[\tilde{\mathcal{R}}+3\,\tilde{\square}\,\text{\ensuremath{\ln}\ensuremath{\omega}}-\frac{3}{2}\left(\tilde{\nabla}\ln\omega\right)^{2}\right],
\end{equation}
\normalsize whereas the term $\sqrt{-g}\,\omega^{2}=\sqrt{-\tilde{g}}$.
The action -- in the Einstein frame -- thus becomes (with the total
derivative $\tilde{\square}$ being dropped) \small
\begin{equation}\label{actionEF}
\int d^{4}x\sqrt{-\tilde{g}}\,\frac{4\Lambda}{\kappa}\left[\tilde{\mathcal{R}}-\frac{3}{2}\left(\tilde{\nabla}\ln\omega\right)^{2}-2\Lambda\right].
\end{equation}
\normalsize Further note that upon variation of action \eqref{eq:action-Einstein-frame}
\begin{equation}
\omega=\frac{\mathcal{R}}{4\Lambda},
\end{equation}

which is equivalent to \[\tilde{\mathcal{R}}=4\Lambda -3\,\tilde{\square}\,\text{\ensuremath{\ln}\ensuremath{\omega}}+\frac{3}{2}\left(\tilde{\nabla}\ln\omega\right)^{2}\,.\]
The metric in the Einstein frame is 
\begin{equation}
\tilde{g}_{\mu\nu}dx^{\mu}dx^{\nu}=-\Psi(R)dt^{2}+\frac{dR^{2}}{\Psi(R)}+r^{2}(R)d\Omega^{2},\label{eq:B2}
\end{equation}
which retains only the proper part of metric \eqref{eq:B1}. This
``seed'' metric, Eq. \eqref{eq:B2}, may be suitable for generalizing
to the stationary axisymmetric setup.

\section{\label{sec:Non-complexification}The non-complexification procedure}

This procedure has been detailed in~\cite{Azreg-Ainou-2014}, we
shall only directly apply it here. The procedure is applicable for
a generic static ``seed'' metric 
\begin{equation}
ds_{\text{stat}}^{2}=-G(R)dt^{2}+\frac{dR^{2}}{F(R)}+H(R)\,d\Omega^{2},\label{s1}
\end{equation}
with $F(R)=G(R)=\Psi(R)=\frac{p(R)q(R)}{r(R)}$, and $H(R)=r^{2}(R)$.
Setting~\cite{Azreg-Ainou-2014} 
\begin{align}
\rho^{2} & :=r^{2}(R)+a^{2}\cos^{2}\theta ,\nonumber \\
2f(R) & :=r^{2}(R)-r(R)p(R)q(R),\nonumber \\
\Delta(R) & :=r(R)p(R)q(R)+a^{2}\label{m-0},\\
\Sigma(R,\theta) & :=[r^{2}(R)+a^{2}]^{2}-a^{2}\Delta(R)\sin^{2}\theta ,\nonumber 
\end{align}
with the evolution rules for $\{p,q,r\}$ given in Eqs. \eqref{eq:evol-1}--\eqref{eq:evol-3}.
A stationary axisymmetric ``candidate'' metric -- \emph{in the
Einstein frame} -- is {\small
\begin{align}
& ds^{2}  =\frac{\psi(R,\theta;a)}{\rho^{2}}\bigg\{-\bigg(1-\frac{2f(R)}{\rho^{2}}\bigg)dt^{2}-\frac{4af(R)\sin^{2}\theta}{\rho^{2}}dt\,d\phi\nonumber \\
\label{eq:ansatz-0}& +\frac{\rho^{2}}{\Delta(R)}dR^{2} +\rho^{2}d\theta^{2}+\frac{\Sigma(R,\theta)}{\rho^{2}}\sin^{2}\theta d\phi^{2}\bigg\},
\end{align}}%
with yet an undetermined $\psi(R,\theta;a)$ satisfying
some partial differential equation. \vskip6pt 

A reverse conformal
mapping would be needed to bring the candidate metric back to the
Jordan frame. Since $\Psi$ is yet determined, perhaps it should be
chosen such that the above metric satisfies the pure $\mathcal{R}^{2}$
vacuo equation. Of course, it is still a daunting task without a guarantee
of eventual success, but it is some progress. It recovers known metrics
as $a\rightarrow{0}$ and/or $k\rightarrow{0}$, and everything looks
gracious in between. \vskip6pt Whatever the reverse conformal mapping
is, the exact rotating metric will be given by the following \emph{ansatz}
{\small
\begin{align}
& ds^{2}=A(R,\theta;a)\Big[-\Big(1-\frac{2f(R)}{\rho^{2}}\Big)dt^{2}-\frac{4af(R)\sin^{2}\theta}{\rho^{2}}dt\,d\phi\nonumber \\
& \ \ \ \ \ +\frac{\rho^{2}}{\Delta(R)}dR^{2}+\rho^{2}d\theta^{2}+\frac{\Sigma(R,\theta)}{\rho^{2}}\sin^{2}\theta d\phi^{2}\Big],\label{eq:ansatz-1}
\end{align}}%
where $A(R,\theta;a)$ satisfies some challenging partial
differential equation that remains to be solved.

\section{\label{sec:Case-of-Lambda0}The case of asymptotically flat $\Lambda=0$}

Let us apply the ansatz \eqref{eq:ansatz-1} to the case $\Lambda=0$
(leading to $q=R$, noting the ``translational invariance'' $R\rightarrow R+R_{0}$),
the static solution of which takes the form ~\cite{Nguyen-2022-Lambda0}
\small
\begin{equation}
ds^{2}=\Big(\frac{q-q_{+}}{q-q_{-}}\Big)^{\frac{k}{q_{+}-q_{-}}}\Big[-\frac{qp(q)}{r(q)}dt^{2}+\frac{r(q)}{qp(q)}dq^{2}+r^{2}(q)d\Omega^{2}\Big]\label{mm2},
\end{equation}
\normalsize where we have used the function $q$ as a radial coordinate
since in this case $R(r)=q(r)$. In this metric we have 
\begin{align}
& q_{+}=\frac{r_{s}}{2}[\sqrt{1+3\tilde{k}^{2}}-1],\quad q_{-}=-\frac{r_{s}}{2}[\sqrt{1+3\tilde{k}^{2}}+1],\nonumber \\
& p(q)=\frac{(q-q_{+})(q-q_{-})}{qr(q)},\quad q_{-}q_{+}=-\frac{3k^{2}}{4},\label{m-3}\\
& r^{2}(q)=(q-q_{+})^{\frac{2q_{+}}{q_{+}-q_{-}}}(q-q_{-})^{\frac{-2q_{-}}{q_{+}-q_{-}}},\nonumber 
\end{align}
where we have introduced the dimensionless parameter $\tilde{k}=k/r_{s}$.\vskip4pt

The rotating solution is given by~\eqref{eq:ansatz-1} with its various
functions as given in~\eqref{m-0} and $R(r)=q(r)$. In particular
\begin{align}
\rho^{2} & :=(q-q_{+})^{\frac{2q_{+}}{q_{+}-q_{-}}}(q-q_{-})^{\frac{-2q_{-}}{q_{+}-q_{-}}}+a^{2}\cos^{2}\theta\nonumber \\
2f(q) & :=(q-q_{+})^{\frac{2q_{+}}{q_{+}-q_{-}}}(q-q_{-})^{\frac{-2q_{-}}{q_{+}-q_{-}}}\nonumber \\
& \ \ \ \ \ -(q-q_{+})(q-q_{-})\nonumber \\
\Delta(q) & :=(q-q_{+})(q-q_{-})+a^{2}\label{m-4}\\
\Sigma(q,\theta) & :=\left[(q-q_{+})^{\frac{2q_{+}}{q_{+}-q_{-}}}(q-q_{-})^{\frac{-2q_{-}}{q_{+}-q_{-}}}+a^{2}\right]^{2}\nonumber \\
& \ \ \ -a^{2}\left[(q-q_{+})(q-q_{-})+a^{2}\right]\sin^{2}\theta\nonumber 
\end{align}
and 
\begin{align}
& ds^{2}=A(q,\theta;a)\Big[-\Big(1-\frac{2f(q)}{\rho^{2}}\Big)dt^{2}-\frac{4af(q)\sin^{2}\theta}{\rho^{2}}dt\,d\phi\nonumber \\
& \ \ \ \ \ +\frac{\rho^{2}}{\Delta(q)}dq^{2}+\rho^{2}d\theta^{2}+\frac{\Sigma(q,\theta)}{\rho^{2}}\sin^{2}\theta d\phi^{2}\Big]\label{m-5}
\end{align}
where the function $A$ satisfies the following nonlinear partial differential equation with $y=\cos\theta$.
\begin{widetext}
{\small
\begin{align*}
&2 \{4 q^3 p^3 r a^2 y^2+q p r [(3 k^2-4 q^2) a^2 y^2+3 k^2 r^2]+4 a^2 q^2 p^2 [a^2 y^2-(1-y^2) r^2]+a^2
[(3 k^2-4 q^2) a^2 y^2+(3 k^2+4 q^2 (1-y^2)) r^2]\} A^2\\
&+6 q^2 p^2 (a^2 y^2+r^2)^2 [(a^2+q
p r) (\partial _q A)^2+(1-y^2) (\partial _y A)^2]-3 q p^2 (a^2 y^2+r^2)^2\\
&\times [(3 k^2+4
q^2+4 q p r) \partial _q A+4 q (a^2+q p r) \partial _{q,q}A+4 q \partial _y ((1-y^2) \partial _yA)]A=0.
\end{align*}}
\end{widetext}
Here $p$ and $r$ are functions of $q$ defined in~\eqref{m-3}. Note that if one sets $a=0$ in this equation, one obtains the differential equation to which the factor $\Big(\dfrac{q-q_{+}}{q-q_{-}}\Big)^{\frac{k}{q_{+}-q_{-}}}$ in~\eqref{mm2} is a solution.\vskip4pt

\begin{figure*}[!htb]
\centering \includegraphics[width=0.44\textwidth]{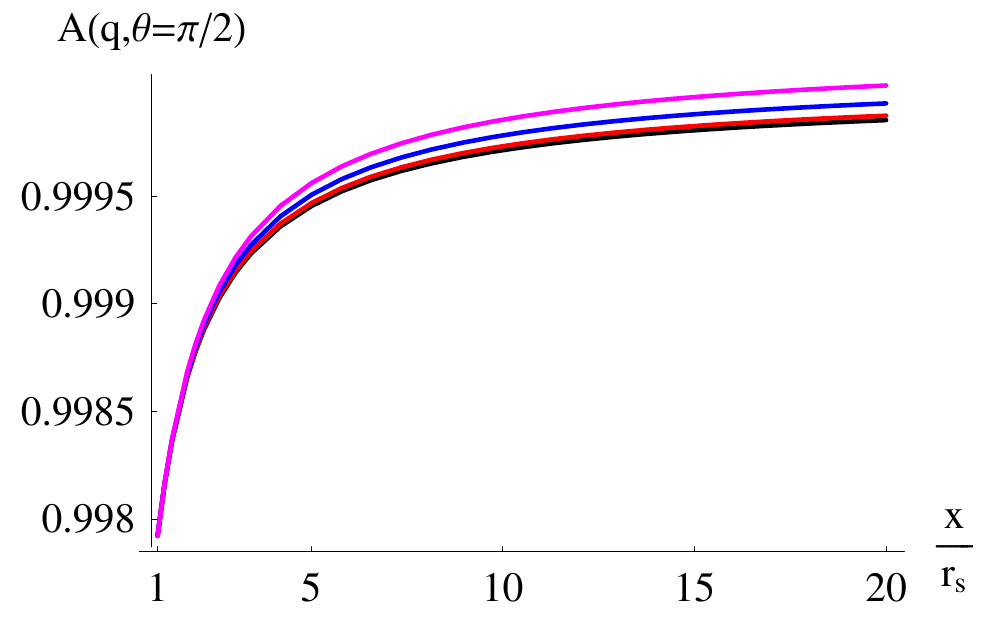}
\centering \includegraphics[width=0.44\textwidth]{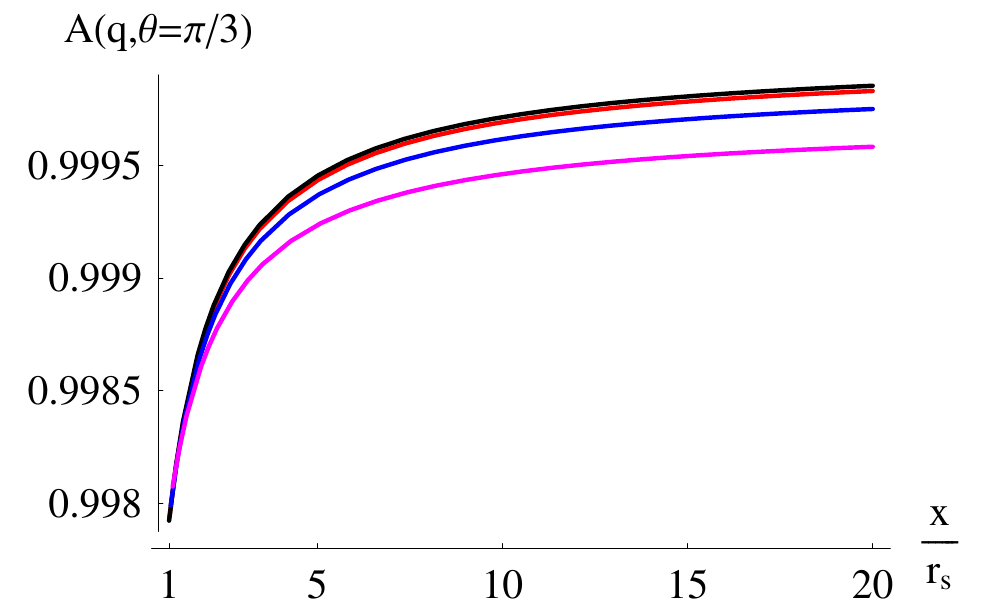}
\caption{\footnotesize{The plots depict the function $A(q,\theta;a)$ versus $x/r_{\text{s}}= \sqrt{q^2+a^2\cos^2\theta}/r_{\text{s}}$ for selected values of ($a/r_\text{s},\,\theta$) and $\tilde{k}=0.003$ has been constrained by the shadow observations~\eqref{ktilde}. The black, red, blue, and magenta curves correspond to $a/r_\text{s}=0,\,0.3,\,0.6,\,0.9$, respectively. The black plot, corresponding to $a=0$, coincides with the plot of $A(q)=\Big(\frac{q-q_{+}}{q-q_{-}}\Big)^{\frac{k}{q_{+}-q_{-}}}$~\eqref{mm2}.}}
\label{FigA1}
\end{figure*}

\begin{figure*}[!htb]
\centering \includegraphics[width=0.44\textwidth]{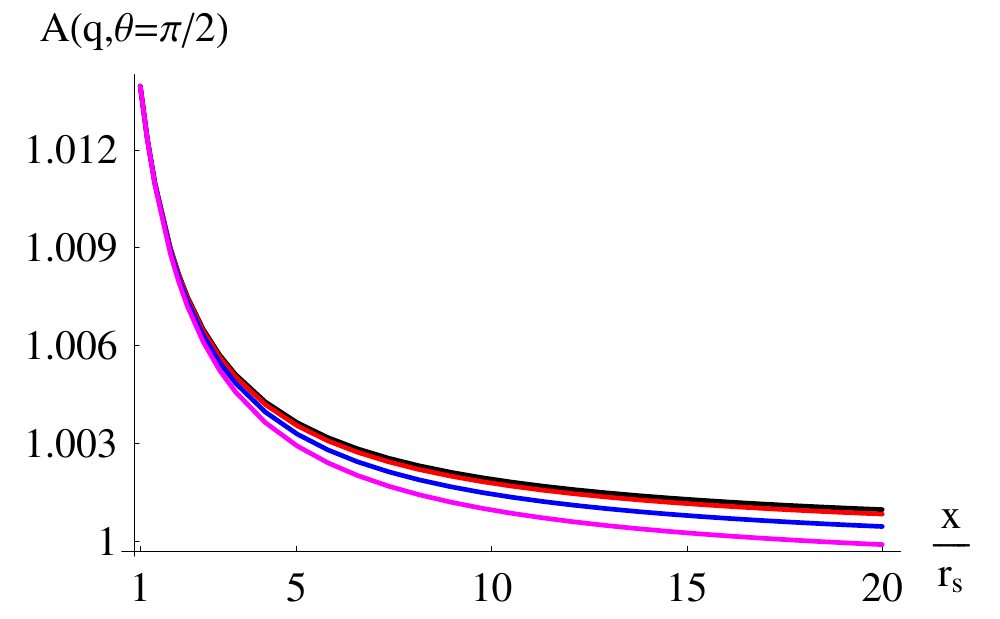}
\centering \includegraphics[width=0.44\textwidth]{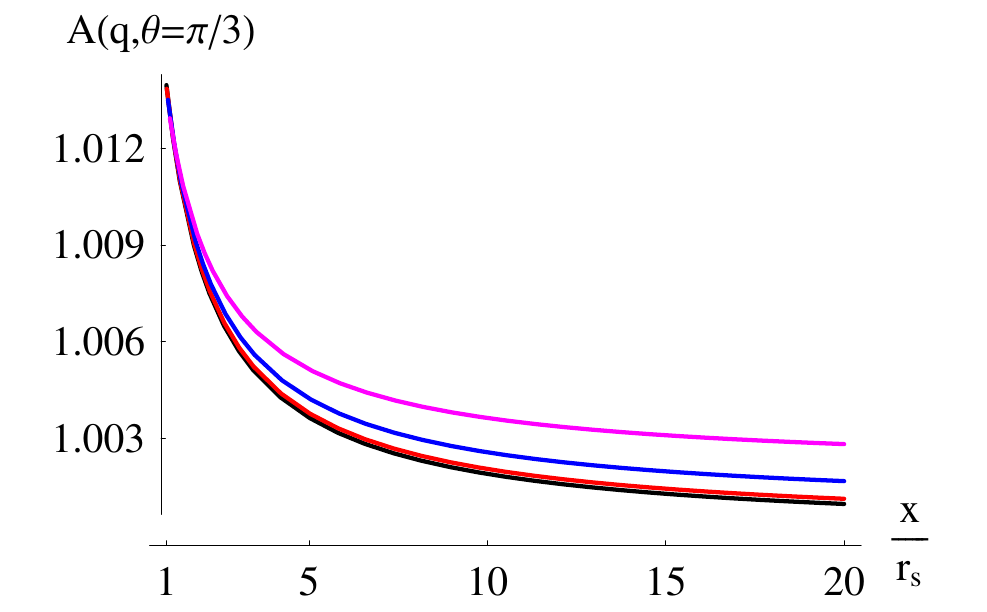}
\caption{\footnotesize{The plots depict the function $A(q,\theta;a)$ versus $x/r_{\text{s}}= \sqrt{q^2+a^2\cos^2\theta}/r_{\text{s}}$ for selected values of ($a/r_\text{s},\,\theta$) and $\tilde{k}=-0.02$ has been constrained by the shadow observations~\eqref{ktilde}. The black, red, blue, and magenta curves correspond to $a/r_\text{s}=0,\,0.3,\,0.6,\,0.9$, respectively. The black plot, corresponding to $a=0$, coincides with the plot of $A(q)=\Big(\frac{q-q_{+}}{q-q_{-}}\Big)^{\frac{k}{q_{+}-q_{-}}}$~\eqref{mm2}.}}
\label{FigA2}
\end{figure*}

The function $A$, as plotted in Figs.~\ref{FigA1} and~\ref{FigA2}, has no zeros, so the horizons are solution to $\Delta(q)=0$ where $\Delta(q)$ is given in Eq.~\eqref{m-4}. Equation $\Delta(q)=0$ reduces to $(q-q_+ )(q-q_- )+a^2=0$, which has at most two horizons and the solutions of which are trivially obtained expressing the new radial coordinate $q$ in terms of the parameters $(q_+,q_-,a)$: $$q=\frac{q_++q_- \pm\sqrt{(q_+-q_-)^2-4a^2}}{2}\,.$$ Here $q_+$ and $q_-$ are defined in Eq.~\eqref{m-3}. The non-rotating case corresponds to $a=0$, yielding $q=q_+$ or $q=q_-$.\vskip4pt

Since in the limit $a\to0$, the metric inside the square brackets in~\eqref{m-5} reduces to the metric inside the square brackets in~\eqref{mm2}, we should take 
\begin{equation}
A(q,\theta;a)=\Big(\frac{q-q_{+}}{q-q_{-}}\Big)^{\frac{k}{q_{+}-q_{-}}}B(q,\theta;a),\label{m6}
\end{equation}
where $\lim_{a\to0}B(q,\theta;a)=1$. Note that we must also have
$\lim_{q\to\infty}B(q,\theta;a)=1$ to preserve asymptotic flatness.
It remains a technical challenge to determine a solution to the partial
differential equation satisfied by the function $B(q,\theta;a)$,
which emanates from that satisfied by the conformal factor $A(R,\theta;a)$~\eqref{eq:ansatz-1}.
However, we are reassured, based on existence theorems, that solutions
to this partial differential equation exist. Numerically we can determine either function, $A(q,\theta;a)$ or $B(q,\theta;a)$, and generate plots for different values of ($a/r_\text{s},\,\theta,\,\tilde{k}$). Figures~\ref{FigA1} and~\ref{FigA2} depict the function $A(q,\theta;a)$ for selected values of ($a/r_\text{s},\,\theta$) and $\tilde{k}$ has been constrained by the shadow observations~\eqref{ktilde}. \vskip4pt

\noindent 

If we identify $B(q,\theta;a)$ with unity we obtain a rotating solution
up to $a^{2}$, that is, $\mathcal{R}=\mathcal{O}(a^{2})+\cdots$
and the left-hand sides of the field equations, \small
\[
\mathcal{R}\Big(\mathcal{R}_{\mu\nu}-\frac{1}{4}g_{\mu\nu}\mathcal{R}\Big)+(g_{\mu\nu}\square-\nabla_{\mu}\nabla_{\nu})\mathcal{R}=0,
\]
\normalsize are of the same order.\vskip4pt

However, for the remaining section of this work we will not identify
$B(q,\theta;a)$ with unity, nor rely on the numerical solutions depicted in Fig.~\ref{FigA1} and~\ref{FigA2}, as for some interesting physical applications
the exact form the conformal factor $A(q,\theta;a)$ is not needed.
This is indeed the case when investigating the shadow of such rotating
solutions as the null geodesic equations are separable for any $A(q,\theta;a)$.

\section{\label{sec:Shadow}Shadow}

In order to investigate the shadow, it is better to rewrite the rotating
metric~\eqref{m-5} keeping the only function $r(q)$ and $\Delta(q)$:
\small
\begin{align}
ds^{2} & =A(q,\theta;a)\Big[-\frac{\Delta-a^{2}\sin^{2}\theta}{\rho^{2}}dt^{2}+\frac{\rho^{2}}{\Delta}dq^{2}+\rho^{2}d\theta^{2}\nonumber \\
& +\frac{2a\sin^{2}\theta}{\rho^{2}}(\Delta-r^{2}-a^{2})dt\,d\phi+\frac{\Sigma}{\rho^{2}}\sin^{2}\theta d\phi^{2}\Big],\label{m-7}
\end{align}
\normalsize where $\rho^{2}(q,\theta)=r^{2}+a^{2}\cos^{2}\theta$
and $\Sigma(q,\theta)=(r^{2}+a^{2})^{2}-\Delta a^{2}\sin^{2}\theta$.\vskip4pt

Separability of the equations of motion for massive and massless particles
has been done for the case where $A(q,\theta;a)\equiv1$ first in~\cite{Azreg-Ainou-2014}
then elsewhere. For the generic case where $A(q,\theta;a)$ is not
constant, the separability of the equations of motion is only possible
for null geodesics (massless particles), as shown in~\cite{Junior}.
The equations describing the null geodesics are straightforwardly
brought to the form
\begin{align}
A^{2}\rho^{4}(\dot{q})^{2}= & [E(r^{2}+a^{2})-aL]^{2}-\Delta[Q+(aE-L)^{2}],\nonumber \\
\equiv & E^{2}\mathbb{R}(q),\label{mtn1}\\
A^{2}\rho^{4}(\dot{\theta})^{2}= & Q+a^{2}E^{2}\cos^{2}\theta-L^{2}\cot^{2}\theta\equiv E^{2}\Theta(\theta),\label{mtn2}\\
\dot{t}= & \frac{E\Sigma}{A\rho^{2}\Delta}+\frac{aL(\Delta-r^{2}-a^{2})}{A\rho^{2}\Delta},\label{mtn3}\\
\dot{\phi}= & \frac{L(\Delta-a^{2}\sin^{2}\theta)}{A\rho^{2}\Delta\sin^{2}\theta}-\frac{aE(\Delta-r^{2}-a^{2})}{A\rho^{2}\Delta},\label{mtn4}
\end{align}
where $E$ and $L$ are the energy and angular momentum, respectively,
$Q$ is the Carter's constant, and the dot denotes derivative with
respect to proper time or affine parameter. In terms of the impact
parameters $\eta\equiv Q/E^{2}$ and $\xi\equiv L/E$, the functions
$\mathbb{R}(q)$ and $\Theta(\theta)$ take the form
\begin{align}
& \mathbb{R}(q)=[r^{2}+a^{2}-a\xi]^{2}-\mathcal{L}\Delta,\label{r}\\
& \Theta(\theta)=\mathcal{L}-(a\sin\theta-\xi\csc\theta)^{2},\label{t}\\
& \mathcal{L}\equiv\eta+(a-\xi)^{2}
\label{L}
\end{align}
Now, we require the presence of unstable spherical null geodesics
obeying the constraints $\mathbb{R}(q)=0$ and $\mathbb{R}'(q)=0$,
where prime denotes derivative with respect to $q$, along with $\mathbb{R}''(q)>0$.
This yields\footnote{The system $\mathbb{R}(q)=0$ and $\mathbb{R}'(q)=0$ has another
solution $\eta=-r^{4}/a^{2}$, $\xi=(r^{2}+a^{2})/a$ yielding $\mathcal{L}=0$.
Now, since $\Theta(\theta)\geq0$, from~\eqref{t} we have $\xi=a\sin^{2}\theta_{0}$
with $\theta_{0}$ being constant. These two expressions of $\xi$
yield $r^{2}+a^{2}\cos^{2}\theta_{0}=0$, which is possible only on
the ring singularity: $r=0,\,\theta_{0}=\pi/2$.}
\begin{align}
& \eta=\frac{r^{2}[8\Delta r'(2a^{2}r'+r\Delta')-16\Delta^{2}(r')^{2}-r^{2}(\Delta')^{2}]}{a^{2}(\Delta')^{2}},\label{eta}\\
& \xi=\frac{r^{2}+a^{2}}{a}-\frac{2\Delta(r^{2})'}{a\Delta'}.\label{xi}
\end{align}
For the determination of the shadow, the celestial coordinates $x$
and $y$, which allow to span the observer's sky, are defined for
a distant observer by \small
\begin{align}
& x=\lim_{r\to\infty}\Big[-r^{2}\sin\theta\frac{d\phi}{dr}\Big|_{\theta=\theta_{i}}\Big]=\lim_{q\to\infty}\Big[-q^{2}\sin\theta\frac{d\phi}{dq}\Big|_{\theta=\theta_{i}}\Big],\\
& y=\lim_{r\to\infty}\Big[r^{2}\frac{d\theta}{dr}\Big|_{\theta=\theta_{i}}\Big]=\lim_{q\to\infty}\Big[q^{2}\frac{d\theta}{dq}\Big|_{\theta=\theta_{i}}\Big],
\end{align}
\normalsize where $\theta_{i}$ is the observer's inclination. Using
Eqs.~\eqref{mtn1}, \eqref{mtn2}, \eqref{mtn4} we obtain: 
\begin{align}
x= & -\xi\csc\theta_{i},\\
y= & \pm\sqrt{\eta+a^{2}\cos^{2}\theta_{i}-\xi^{2}\cot^{2}\theta_{i}}\,,\nonumber \\
= & \pm\sqrt{\mathcal{L}-(a\sin\theta_{i}-\xi\csc\theta_{i})^{2}}\,.
\end{align}
These last equations along with~\eqref{eta} and~\eqref{xi} will
allow us to sketch the shape of the shadow angular size versus the
dimensionless parameter $\tilde{k}=k/r_{s}$. From the static solution~\eqref{mm2}
we obtain 
\begin{equation}
r_{s}=\frac{2M}{1+\tilde{k}},\label{rs}
\end{equation}
which will allow us to express the functions of the rotating solution~\eqref{m-3}
and~\eqref{m-4} in terms of $M$, $\tilde{k}$, and $a_{*}=a/M$.
\vskip4pt
\begin{figure}[!htb]
\centering \includegraphics[width=0.47\textwidth]{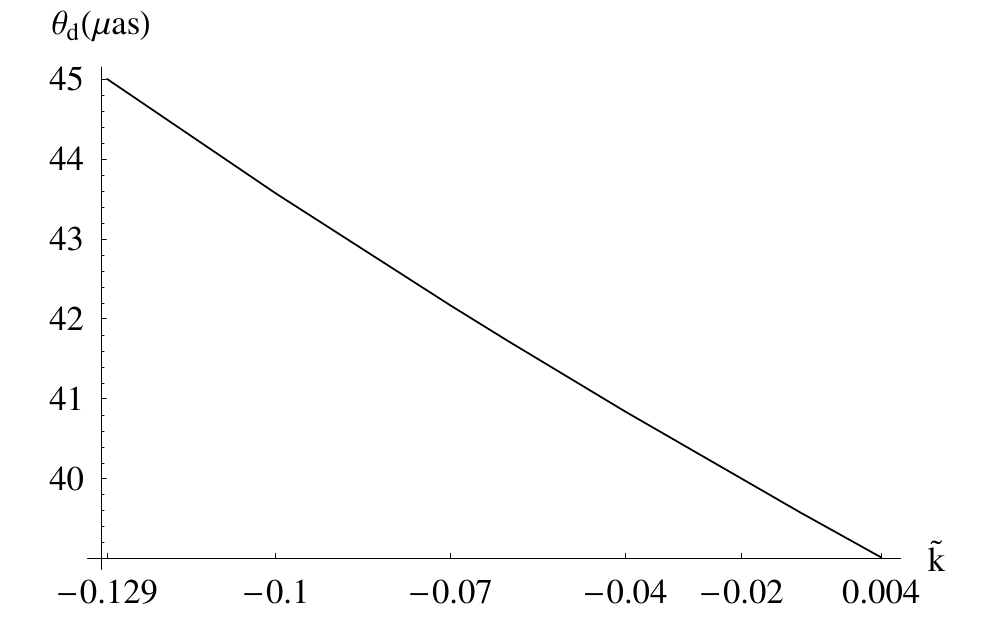}
\caption{{\footnotesize{}{The plot depicts the shadow angular size $\theta_{d}$
			versus the dimensionless parameter $\tilde{k}$ assuming that M87{*}
			if modeled by the rotating metric~\eqref{m-7}. If $\tilde{k}=k/r_{s}\in[-0.129,\,0.004]$,
			then $\theta_{d}=42\pm3\,\mu\text{as}$~\cite{EHT1,EHT2,EHT3,EHT4}.
			We took $M=6.5\times10^{9}M_{\odot}$, $a_{*}\equiv a/M=0.5$, distance
			to Earth $D=16.8\,\text{Mpc}$, inclination $\theta_{i}=17^{\text{o}}$.}}}
\label{Fig1}
\end{figure}

\noindent 
\begin{figure}[!htb]
\centering \includegraphics[width=0.47\textwidth]{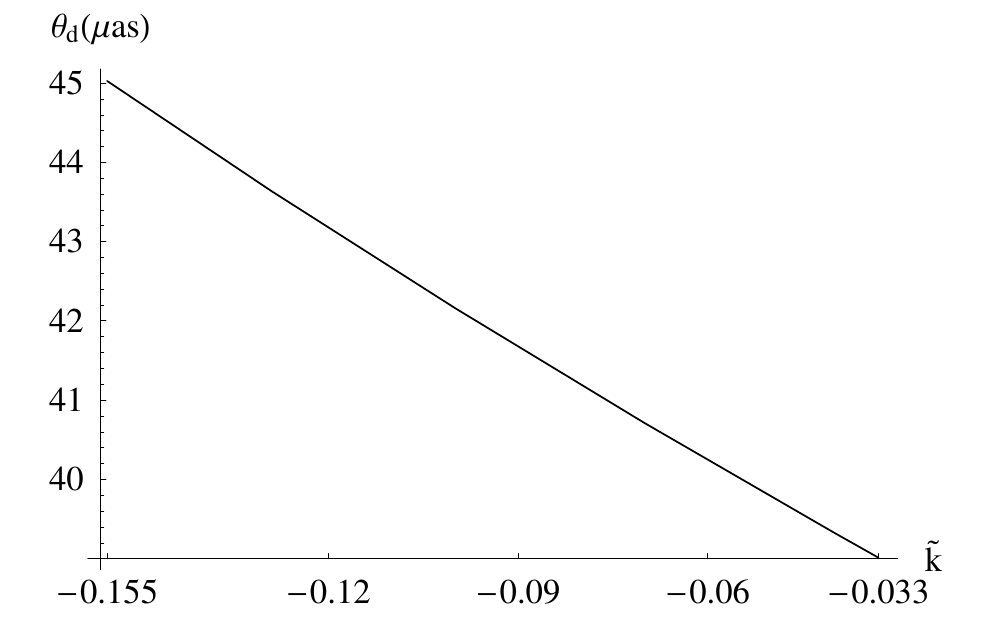}
\caption{{\footnotesize{}{The plot depicts the shadow angular size $\theta_{d}$
			versus the dimensionless parameter $\tilde{k}$ assuming that M87{*}
			if modeled by the rotating metric~\eqref{m-7}. If $\tilde{k}=k/r_{s}\in[-0.155,\,-0.033]$,
			then $\theta_{d}=42\pm3\,\mu\text{as}$~\cite{EHT1,EHT2,EHT3,EHT4}.
			We took $M=6.5\times10^{9}M_{\odot}$, $a_{*}\equiv a/M=0.94$, distance
			to Earth $D=16.8\,\text{Mpc}$, inclination $\theta_{i}=17^{\text{o}}$.}}}
\label{Fig2}
\end{figure}

The shadow angular size $\theta_{d}$ is defined by 
\begin{equation}
\theta_{d}=\frac{2R_{s}}{D},\label{key}
\end{equation}
where $D$ is the distance to Earth and $R_{s}$ is the radius of
the circle that shares with the closed curve describing the shadow
the following three points: The rightmost point of the shadow ($x=x_{r},\,y=0$),
the upper- and lowermost points of the shadow ($x=x_{b},\,y=y_{\text{max}}$)
and ($x=x_{b},\,y=-y_{\text{max}}$). These are the three red points
of Figure~3 of Ref.~\cite{Maeda} where $\alpha$ denotes $x$
and $\beta$ denotes $y$: 
\begin{equation}
R_{s}=\frac{(x_{r}-x_{b})^{2}+y_{\text{max}}^{2}}{2(x_{r}-x_{b})}.\label{Rs}
\end{equation}
\vskip4pt

Now, if the rotating metric~\eqref{m-7} describes the central black
hole M87{*} we must have $M=6.5\times10^{9}M_{\odot}$, $a_{*}\equiv a/M\in[0.5,0.94]$,
distance to Earth $D=16.8\,\text{Mpc}$, inclination $\theta_{i}=17^{\text{o}}$,
and shadow angular size $\theta_{d}=42\pm3\,\mu\text{as}$~\cite{EHT1,EHT2,EHT3,EHT4}.
This yields the bounds for $\tilde{k}\in[-0.129,\,0.004]$ if $a_{*}=0.5$
and $\tilde{k}\in[-0.155,\,-0.033]$ if $a_{*}=0.94$, as shown in
Fig.~\ref{Fig1} and Fig.~\ref{Fig2}, respectively. These two bounds
restrict $\tilde{k}$ by 
\begin{equation}
-0.155\leq\tilde{k}\leq 0.004\,.\label{ktilde}
\end{equation}

\section{\label{sec:Stability}Stability analysis}

Upon comparing our action in the Einstein frame~\eqref{actionEF} with that in the same frame given in Eq.~(2.10) of~\cite{BRS23}, we make the following identification
\begin{equation}\label{sa1}
M^2=\frac{8\Lambda}{\kappa},\quad \zeta =\sqrt{\frac{12\Lambda}{\kappa}}\,\ln\omega,\quad U(\zeta)=\frac{8\Lambda^2}{\kappa},
\end{equation}
and $\rho\equiv 0$. If ($h_{\mu\nu},\,\delta\zeta$) denote the perturbations of the fields,
\begin{equation*}
g_{\mu\nu}=\tilde{g}_{\mu\nu}+h_{\mu\nu},\qquad \zeta=\tilde{\zeta}+\delta\zeta ,
\end{equation*}
where the tilde notation is used for the background fields in the Einstein frame, the field equations reduce to~\cite{BRS23}
\begin{equation}\label{sa2}
\delta \tilde{G}_{\mu\nu}+\Lambda h_{\mu\nu}=0,\qquad \tilde{\square}(\delta\zeta) =0 .
\end{equation}
Since these are the same perturbation equations of general relativity in a de Sitter background in the presence of a scalar field which has led to the stability of the static Schwarzschild--de Sitter black hole against linear perturbations~\cite{BRS20}, we conclude that our static solutions, regardless of the value of $\tilde{\mathcal{R}}$ ($\tilde{\mathcal{R}}=0$ or $\tilde{\mathcal{R}}\neq 0$), are also stable against linear perturbations. We could have reached the same conclusion upon applying the analysis given either in~\cite{BRS20,Zerilli} or in~\cite{s4}.\vskip4pt

Since the conformal factor is not given analytically, the stability of rotating solutions cannot be performed following the work done in~\cite{BRS20} or any other reference. However, for small rotation parameter $a$ we may claim stability of rotating solutions by continuity.\vskip4pt

We conclude that the non-rotating solution is stable against linear radial perturbations and we claim that the rotating solution is also stable as far as the rotating parameter remains small compared to the mass of the solution.

\section{\label{sec:Summary}Conclusion}

In this work we have emphasized the role of the Buchdahl parameter $k$, which has dimensions of length
and is of higher-derivative nature. We have shown that the Buchdahl parameter $k$ can be considered as the scalar charge and should take on any in real value. 

Upon making a recap of the general and special non-rotating Buchdahl-inspired metrics, we proceeded to transform the general non-rotating Buchdahl-inspired metric to a seed metric suitable for generalizing its rotating counterpart. Then we applied the non-complexification procedure of the Newman-Janis algorithm to reduce the number of unknown functions of the rotating metric to 1: This is the function $\psi$ in Eq.~\eqref{eq:ansatz-0} or the function $A$ in Eq.~\eqref{eq:ansatz-1}. 

After that, we specialized to the special metric having $\Lambda =0$ and obtained its exact rotating counterpart, Eqs.~\eqref{m-4} \& \eqref{m-5}, up to a conformal factor $A$, which we could not determine analytically but numerically, upon solving the equation $\mathcal{R}=0$, as depicted in Figs.~\ref{FigA1} \& \ref{FigA2}. 

The shadow analysis of the exact rotating solution has shown that the reduced Buchdahl parameter $\tilde{k}$ could be restricted by $-0.155\leq\tilde{k}\leq 0.004$~\eqref{ktilde}.

Finally, we concluded that the non-rotating solution is stable against linear radial perturbations. The stability of the rotating solution cannot be performed as $A$ is not given analytically; however, we may claim that it is continuous too by continuity at least for small values of the rotation parameter compared to the mass of the solution.

\begin{acknowledgments}
We thank Richard Shurtleff for his technical help during the development
of this work, and Tiberiu Harko for his fruitful suggestion of making
use of the Einstein frame in place of the Jordan frame.
\end{acknowledgments}

\end{document}